\documentclass[twocolumn,preprintnumbers,amssymb,amsmath,superscriptaddress,letterpaper,nofootinbib]{revtex4}[12pt]

\usepackage{graphicx}
\usepackage{dcolumn}
\usepackage{bm}
\usepackage{natbib}
\usepackage{pstricks}
\usepackage{epsfig}
\usepackage{epstopdf}

\usepackage{mathbbol}

\newcommand{\be}{\begin{equation}}
\newcommand{\ee}{\end{equation}}
\newcommand{\bea}{\begin{eqnarray}}
\newcommand{\eea}{\end{eqnarray}}

\begin{document}

\title{Path Integrated Geodesics and Distances}

\author{Nima Khosravi}
\email{nima@sharif.edu}
\affiliation{Department of Physics, Sharif University of Technology, Tehran 11155-9161, Iran}

\date{\today}

\begin{abstract}
 In this paper, the quantum corrections to the kinematics of geometry, specifically geodesics, are presented. This is done by employing the path integral over the geodesics. Interestingly, the geodesics do not see any modifications in this framework. However for the distances,  it is demonstrated that these quantum corrections exhibit distinct behaviors for time-like, light-like, and space-like geodesics. For time-like geodesics, the maximum correction is the Planck length, which disappears when the classical separation vanishes. The light-like geodesics do not exhibit quantum corrections, meaning that the causal light cone remains the same in both classical and quantum frameworks under certain conditions. The quantum corrections for space-like geodesics impose a minimum on space-like separation, potentially playing a role in removing singularities by preventing null congruences from being closer than the Planck length. This framework also explores the correspondence between space-like/time-like geodesics and quantum/statistical physics.

\end{abstract}
\maketitle

\section{Introduction:}
The quest for the quantum aspect of gravity (and geometry) has become one of the most important areas in theoretical physics, if not in the entire physics community. The two pillars of theoretical physics—quantum theory and Einstein's gravity—have successfully described almost all observations and experiments. Moreover, both theories have profoundly changed our conceptual understanding of how nature works. However, there are a few crucial observations that cannot be explained by our current theories, such as the singularities of the Big Bang and black holes, as well as the hierarchy problems.

Einstein taught us that the gravitational field can be viewed as the geometrical curvature of spacetime. In this framework, the metric $g_{\mu\nu}$ plays a central role, providing distances through $ds^2=g_{\mu\nu}dx^\mu dx^\nu$. At the level of dynamics, the Einstein-Hilbert action governs everything. Here, we focus on the kinematics of a curved geometry, where the geodesic equation is the main component. In principle, the geodesic equation can be derived by knowing the connection, which can be interpreted as a new geometrical object independent of the metric. We assume the connection is the Levi-Civita symbol and is given by the metric. In this case, the geodesic equation can be deduced from the least action principle, with the Lagrangian given by:
\begin{eqnarray}\label{c-geo}
{\cal{S}}_C&\equiv&\int ds= \int_A^B\,\sqrt{g_{\mu\nu}dx^\mu dx^\nu} \\ \nonumber &=&\int_A^B\,d\lambda L_C
\end{eqnarray}
where $L_C\equiv \sqrt{g_{\mu\nu}\frac{dx^\mu}{d\lambda} \frac{dx^\nu}{d\lambda}}$ and $\lambda$ is an auxiliary parameter.The Euler-Lagrange equations for the variation of ${\cal{S}}_C$ with respect to $g_{\mu\nu}$ result in the geodesic equation. From this viewpoint, the geodesic equation represents an extremum in the space of all possible paths between points A and B in spacetime. The subscript $C$ in ${\cal{S}}_C$ denotes "classical." Assuming the $(-,+,+,+)$ signature, the absolute distances ($|ds^2|$) are maximized and minimized for time-like and space-like geodesics, respectively.

In this work, we explore the quantum nature of geometry through the following idea. One way to realize the quantization procedure is via the Feynman path integral, which sums over all possible paths, each with a specific probability. What if we apply a similar approach to geodesics?

\section{Integration over geodesics:}
We begin with modifying the relation (\ref{c-geo}) by
\begin{eqnarray}
{\cal{S}}_Q\equiv\int_A^B\,d\lambda L_Q
\end{eqnarray}
where $L_Q\equiv L_C e^{-\beta L_C}$. Now the corresponding Euler-Lagrange equation will be $\dot{p}_\mu=F_\mu$ where
\begin{eqnarray}\nonumber
p_\mu&\equiv&\frac{\partial L_Q}{\partial \dot{x}^\mu}=(1-\beta L_C)\,e^{-\beta\,L_C}\frac{1}{L_C}g_{\mu\alpha}\dot{x}^\alpha,\\\nonumber
F_\mu&\equiv&\frac{\partial L_Q}{\partial {x}^\mu}=(1-\beta L_C)\,e^{-\beta\,L_C}\frac{1}{2 L_C}g_{\alpha\gamma,\mu}\dot{x}^\alpha\dot{x}^\gamma.
\end{eqnarray}
According to the above relations, the Euler-Lagrange equation will take the following form
\begin{eqnarray}\nonumber
\ddot{x}^\alpha+\Gamma^\alpha_{\mu\nu}\dot{x}^\mu\dot{x}^\nu=\dot{x}^\alpha\frac{\dot{L_C}}{L_C}\bigg[1+\beta \frac{L_C(2-\beta L_C)}{1-\beta L_C} \bigg],
\end{eqnarray}
where $\Gamma^\alpha_{\mu\nu}$ is metric connection. The second term in the right hand side is the quantum correction and vanishes if $\beta\rightarrow0$. The first term in right hand side is also well-known and is related to our parametrization. If one choose the affine parameter then $\dot{L_C}$ vanishes and the standard geodesic form appears. But very interestingly, the quantum correction term is also disappears with this choice. This means that the quantum correction, as we proposed, does not modify the geodesic equation. Note that $\dot{L_C}=0$ implies $\dot{L_Q}=0$.

To discuss more about this result, let's remind that it is well-known that the (canonical) geodesic equation can be found by introducing the 
\begin{eqnarray}\label{c2-geo}
_2{\cal{S}}_C&\equiv&\int ds= \int_A^B\,{g_{\mu\nu}dx^\mu dx^\nu} \\ \nonumber &=&\int_A^B\,d\lambda\,\, L^{(2)}_C
\end{eqnarray}
instead of (\ref{c-geo}). In this formulation the Euler-Lagrange variation of $L^{(2)}_C$  reaches to
\begin{eqnarray}\nonumber
\ddot{x}^\alpha+\Gamma^\alpha_{\mu\nu}\dot{x}^\mu\dot{x}^\nu=0,
\end{eqnarray}
without any additional term. This means the geodesic equation is parametrized by an affine parameter, automatically. Now we take the above Lagrangian as the classical one and do the same recipe to get its quantum version. The quantum Lagrangian becomes $L^{(2)}_Q\equiv L^{(2)}_C e^{-\beta L^{(2)}_C}$ and the corresponding Euler-Lagrange equation of motion is
\begin{eqnarray}\nonumber
\ddot{x}^\alpha+\Gamma^\alpha_{\mu\nu}\dot{x}^\mu\dot{x}^\nu=\beta \dot{x}^\alpha \, \frac{dL^{(2)}_C}{dt}\,\bigg[ \frac{2-\beta\, L^{(2)}_C}{1-\beta\, L^{(2)}_C} \bigg].
\end{eqnarray}
The above result clearly shows that the quantum effects show themselves in non-affineness of geodesic equation. As mentioned above, this (quantum) correction term can be removed by changing the arbitrary parameter to an affine parameter. Though in this new notation we can see what is an affine parameter for the classical path is not the one for the quantum path.

However, we can view this result from another perspective, which we would like to discuss. What is an affine parameter? It is the parameter for which the geodesic equation takes its standard form. But from the standard viewpoint, we believe the physics is unchanged if we use other parameters. If we accept this viewpoint, then we cannot observe quantum effects by looking only at the geodesic paths. Yet, if we assume that quantum effects should be observable within the geodesic paths themselves, then we must accept that not all parametrizations of geodesic paths are equivalent. This is a radical viewpoint. It can be seen as equivalent to positing a universal time (or, more generally, length) unit. If a universal time or length unit exists, then we cannot freely change the parametrization as we do in classical cases. Such a universal length scale becomes accessible if one considers modifications to Lorentz symmetry—for example, Doubly Special Relativity, which has its roots in the concept of quantum spacetime  \cite{Kowalski-Glikman:2004fsz}.

\section{Integration over distances:}
Inspired by Feynman's path integral quantization scheme, we perform the integration over geodesics between two given points A and B. The key idea in the path integral formulation is the probability assigned to each path, given by $e^{-i s / l_p}$, where $s$ is the path length and $l_p$ is a constant with length dimension. Therefore, the expectation value for the distance between points A and B is given by:
\begin{eqnarray}
{\cal{S}}_Q=\langle{\cal{S}}\rangle= \frac{\int_A^B\,ds\,s\,e^{-i\,s/l_p}}{\int_A^B\,ds\,e^{-i\,s/l_p}},
\end{eqnarray}
where the subscript $Q$ represents ``quantum".

In our proposal, we restrict ourselves to a condition: if the classical geodesic\footnote{By classical geodesic between two points, we mean the standard geodesic in general relativity.} between points A and B is space-like, null, or time-like, then we sum over all space-like, null, or time-like paths between A and B, respectively\footnote{We believe this condition can be relaxed in future works.}.

\subsection{space-like geodesics}
This part contains perhaps the most interesting result and possibly the main one. We perform an integral over all possible space-like paths between two points A and B. Since being a space-like path is invariant under coordinate transformations, we assume the following integral:
\begin{eqnarray}
{\cal{S}}^{space}_Q= \bigg|\frac{\int_A^B\,ds\,s\,e^{-i\,s/l_p}}{\int_A^B\,ds\,e^{-i\,s/l_p}}\bigg|
\end{eqnarray}
where $l_p$ is a constant with length dimension and can be interpreted as analogous to $\hbar$ in the path integral formalism. Now, we aim to study the quantum corrections to classical space-like geodesics. Note that for space-like separations, the integral can be complex, and to obtain a real value, we need to take the absolute value. Let's consider two space-like separated points A and B with a classical distance $l_C$. Since we know this length is the minimum among all space-like paths between A and B, we can set the lower bound of the integral to $l_C$. On the other hand, the largest space-like separation between two arbitrary points can be infinity, which sets the upper bound of the integral. Therefore, we can express the quantum space-like separation as:
\begin{eqnarray}
{\cal{S}}^{space}_Q= \bigg|\frac{\int_{l_C}^\infty\,ds\,s\,e^{-i\,s/l_p}}{\int_{l_C}^\infty\,ds\,e^{-i\,s/l_p}}\bigg|.
\end{eqnarray}
We can calculate this integral\footnote{To be precise, we need to add a regulator to the integrand as $e^{-\epsilon s}$ and, after calculating the integral, take the limit $\epsilon\rightarrow 0$.}$^,$\footnote{Note that for mathematical rigor in 3-dimensional space, we may need to modify the probability factor to  $e^{-i\,\beta\,s/\l_p}$. This modification helps avoid divergence caused by the increasing number of paths for a given length  $s$.}  and the result will be:
\begin{eqnarray}
l_Q^2= l_C^2+l_p^2
\end{eqnarray}
where $l_Q={\cal{S}}^{space}_Q$. Obviously, when $l_C\gg l_p$, the quantum corrections are negligible. The opposite limit is more interesting:  the minimum quantum distance between two points (in the case where $l_C=0$) is $Min(l_Q)=l_p$, which differs from the classical case, where it is zero. This can have a crucial application in singularity theorems. The separation between, for example, null geodesics is space-like (on a time-constant hypersurface), and according to our results, quantum corrections do not allow this separation to be zero. This is a crucial property for the validity of singularity theorems. Therefore, we can claim that there are no singularities in the presence of quantum corrections. Although this requires more careful analysis, it is not far from the expectations predicted by some quantum gravity models \cite{Bojowald:2001xe,Mathur:2005zp}.

Based on the above results, we can calculate the quantum corrections to areas and volumes. One might easily assume that these quantities are simply the square and cube of the length, respectively. However, another approach is to calculate these values from scratch. For the area, this can be done as follows:
\begin{eqnarray}
{\cal{A}}^{space}_Q= \bigg|\frac{\int_{l_C}^\infty\,ds\,s^2\,e^{-i\,s/l_p}}{\int_{l_C}^\infty\,ds\,e^{-i\,s/l_p}}\bigg|=\sqrt{l_C^4+4\,l_p^4}
\end{eqnarray}
and for the volume
\begin{eqnarray}
{\cal{V}}^{space}_Q= \bigg|\frac{\int_{l_C}^\infty\,ds\,s^3\,e^{-i\,s/l_p}}{\int_{l_C}^\infty\,ds\,e^{-i\,s/l_p}}\bigg|=\sqrt{l_C^6-3\,l_C^4\,l_p^2+36\,l_p^6}
\end{eqnarray}
where both quantum corrections to areas and volumes reduce to the classical results when  $l_C\gg l_p$, but they exhibit minimum values for the case of  $l_C=0$: $Min({\cal{A}}^{space}_Q)=2\,l_p^2$ and $Min({\cal{V}}^{space}_Q)=6\,l_p^3$.

\subsection{light-like geodesics}
In our proposal, the averaging over light-like paths is trivial. Since all these paths are light-like, the path length $s=0$ and consequently, the quantum geodesic also vanishes. This implies that quantum corrections do not contribute to the null cone. This result is interesting because it indicates that the causal structure remains unmodified by quantum corrections.

\subsection{time-like geodesics}
In this case the geodesic distance is purely imaginary, $s=i\,\tau=i\sqrt{g_{\mu\nu}dx^\mu dx^\nu}$, where $\tau$ is a positive real quantity. The quantum distance will be
\begin{eqnarray}
{\cal{S}}^{time}_Q= \frac{\int_A^B\,d\tau\,\tau\,e^{+\tau/l_p}}{\int_A^B\,d\tau\,e^{+\tau/l_p}}.
\end{eqnarray}
The probability factors for each path are very similar to the Boltzmann factor in statistical physics. However, there is a crucial issue that can make it ill-defined: the exponent is positive instead of negative in the Boltzmann factor. This may cause divergence in the integrals. Nevertheless, there is an essential observation regarding timelike geodesics that can solve this problem: the classical timelike geodesic between two points A and B is the \textit{maximum} among all timelike paths connecting these points. This implies that there is an upper limit for the integrals, and they are finite. For a one-dimensional timelike path, the integral can be written as
\begin{eqnarray}
{\cal{S}}^{time}_Q= \frac{\int_0^T\,d\tau\,\tau\,e^{+\tau/l_p}}{\int_0^T\,d\tau\,e^{+\tau/l_p}}=T\frac{1}{1-e^{-T/l_p}} -l_p
\end{eqnarray}
where we assume A is at $0$ and B is at $T$ without loss generality. For very large classical separations i.e. $T\gg l_p$ the quantum distance is very close to its classical value, up to a correction which is smaller than $l_p$. In this case, the minimum value of quantum timelike separations is zero; there is no nontrivial minimum timelike separation. This is understandable since the classical timelike geodesic is the maximum among all timelike paths. It means that the average over all timelike paths is always less than the classical value. Therefore, when the classical separation approaches zero, the quantum one also approaches zero. This is different from the spacelike case.

\section{The ``quantum/statistical physics" and the ``space-like/time-like" duality}
It is evident that averaging over spacelike geodesics is analogous to Feynman's path integral procedure for quantization. However, averaging over timelike geodesics introduces new considerations. To explore the correspondence between quantum timelike geodesics and thermodynamics, let's consider a discrete version without loss of generality. We begin by examining an expectation value as
\begin{eqnarray}
\langle E \rangle=\frac{\sum_{i=0}^{N} E_i e^{+\beta E_i}}{\sum_{j=0}^{N} e^{+\beta E_j}}
\end{eqnarray}
where $\beta$ is a positive fixed parameter and $E_N$ is the maximum of $E_i$'s. Now, we can rewrite the above relation as follow
\begin{eqnarray}\label{duality}\nonumber
\langle E \rangle&=&\frac{\sum_{i=0}^{N} E_i e^{+\beta E_i}}{\sum_{j=0}^{N} e^{+\beta E_j}}\times\frac{e^{-\beta E_N}}{e^{-\beta E_N}}=\frac{\sum_{i=0}^{N} E_i e^{-\beta (E_N-E_i)}}{\sum_{j=0}^{N} e^{-\beta (E_N-E_j)}}\\\nonumber
&=&\frac{\sum_{i=0}^{N} (E_N-\hat{E}_i) e^{-\beta \hat{E}_i}}{\sum_{j=0}^{N} e^{-\beta \hat{E}_j}}=E_N-\frac{\sum_{i=0}^{N} \hat{E}_i e^{-\beta \hat{E}_i}}{\sum_{j=0}^{N} e^{-\beta \hat{E}_j}}\\
&=&E_N-\langle \hat{E} \rangle_{Boltzmann}
\end{eqnarray}
where $\hat{E}_i \equiv E_N -E_i$ and $\langle ..\rangle_{Boltzmann}$ is the standard thermodynamic expectation value with Boltzmann factors. Thus, statistical physics and quantum timelike geodesics obey the same rules up to the aforementioned duality.

\section{Conclusions}
In this paper, we demonstrated that applying the path integral formalism to geometrical geodesics yields several interesting consequences. The first observation is the equivalence of the classical and quantum geodesics. The quantum effects show themselves in the parametrization of the paths which make them unobservable in the standard viewpoint. In addition we studied the quantum effects on the distances between two space/null/time-like points. One of the most significant outcomes is the prediction of a non-vanishing minimum separation for space-like separations. The concept of a minimum length and its implications have been explored in the literature (e.g., see \cite{Hossenfelder:2012jw} and references therein), although in different contexts and frameworks. Another intriguing aspect of this idea is the correspondence between space-like/time-like geodesics and quantum/statistical physics. This correspondence is interesting because it may be rooted in the ``non-local"/``local" structure of these areas of physics. Thermodynamics, based on classical physics, involves very local interactions, which share a similar structure with time-like geodesics, as we have shown. However, space-like geodesics, which connect non-local points, exhibit relationships with quantum physics, which itself possesses non-local structures. This observation requires further elaboration and contemplation. The correspondence between thermodynamics and quantum mechanics can be observed via the Wick rotation, $t\rightarrow - i \beta$, where $1/\beta$ plays the role of temperature in thermodynamics. However, our proposal suggests (perhaps) a crucial fact: the Wick rotation cannot fully capture this correspondence unless we consider (\ref{duality}). This implies that to establish the correspondence, we need a maximum energy state. The existence of this maximum energy is not unphysical, as we know all systems, including our universe, are finite. On the other hand, having a maximum energy is consistent with having finite resolution in measuring distances. The proposed framework, averaging over geodesics, can be interpreted as averaging over connections (and consequently metrics), a concept introduced in \cite{Khosravi:2013kha,Khosravi:2014mua} with other applications. 

\section*{Acknowledgement}
I would like to thank Martin Bojowald for his comment on this manuscript.

\end{document}